\documentclass{amsart}

\newtheorem{theorem}{Theorem}[section]

\theoremstyle{definition}

\theoremstyle{remark}
\newtheorem{remark}[theorem]{Remark}

\numberwithin{equation}{section}

\begin{document}
\title{On S-duality and Gauss reciprocity law}
\author{An Huang}
\address{Department of Mathematics, UC Berkeley}
\email{anhuang@berkeley.edu}
\date{October 08, 2009}

\begin{abstract}
\noindent We review the interpretation of Tate's thesis by a sort of conformal field theory on a number field in \cite{1}. Based on this and the existence of a hypothetical 3-dimensional gauge theory, we give a physical interpretation of the Gauss quadratic reciprocity law by a sort of S-duality. 
\end{abstract}
\maketitle
\section*{Introduction}
The Gauss quadratic reciprocity law is arguably one of the most famous theorems in mathematics. Trying to generalize it from different directions has been a central topic in number theory for centuries. In the form of Artin reciprocity law, or class field theory, abelian reciprocity laws are well-understood and has been a beautiful part of algebraic number theory for a long time. However, the sought for general reciprocity laws is still in a very unclear stage, for example the Langlands program, regarded as a nonabelian generalization of class field theory, is far from settled. 

On the other hand, S-duality (or strong-weak duality) is a common name for many amazing stories in physics, which can be traced back at least to the theoretical work on electric-magnetic duality and magnetic monopoles. S-duality for classical or quantum $U(1)$ gauge theory is more or less well-understood theoretically. However, S-duality for nonabelian gauge theories remains mysterious, or at least largely conjectural. In recent years, people are making huge progress in interpreting the geometric Langlands program by nonabelian S-dualities. One may consult \cite{Witten3} for such an example.

However, as far as we know, there is little work on trying to interpret number theory reciprocity laws by S-duality of some physical theory. One of the reasons is, of course, gauge theories are geometrical in nature, so it seems not natural to try to directly relate S-dualities with reciprocity laws. However, as we mentioned above, the geometric Langlands program, which is the geometric counterpart of the number theory Langlands program, now has good physical interpretations. So we hope, at least, that one can find some sort of physical interpretations of abelian reciprocity laws in number theory. What we will try to do in this paper, is to show that a sort of hypothetical abelian S-duality gives rise to the quadratic reciprocity law with possible generalizations to some higher power reciprocity laws. Of course, because number theory has a different nature, in order to do this in a somewhat direct way, probably one has to leave aside most existing geometric stories, but only to keep in mind basic principles of quantum field theory, and to make use of the geometric picture of number theory to take analogues. Then, one can try to get some sort of physics models to describe some number theory. This is the philosophy of discussion in this paper.

In section 1, we review section 5 of \cite{1}, which is an interpretation of Tate's thesis by a sort of conformal field theory on a number field. One may consult \cite{1} for it, but we will include the discussion here for convenience. Moreover, we will get a better understanding of this reinterpretation in the present paper by considerations coming from the 3-dimensional theory. This 2-dimensional conformal field theory on a number field, will be used as the central tool for us to get a physical interpretation of the quadratic reciprocity law.

In \cite{Witten}, Witten proposed to study some reciprocity laws for function fields by studying quantum field theories on algebraic curves. (One can see remarks at the end of section IV in \cite{Witten}) This is one of the sources of our ideas. Moreover, we will use ideas from \cite{Witten2}, and assume that our 2-dimensional theory is originated from a $\operatorname{GL}(1)$ 'gauge theory' living on $\texttt{Spec} \textsl{O}_K$, or the number field $K$, regarded as 3-dimensional in the point of view of the etale cohomology, which is expected to be the arithmetic counterpart of 3-dimensional Chern-Simons theory with $G=U(1)$ with some sort of S-duality. Then this S-duality should reflect itself in some natural way in the path integral of our 2-dimensional theory living on the same number field regarded as 2-dimensional, as we will see. (In physics terminology, probably we should name the above description as a form of AdS/CFT correspondence. But we will not try to discuss the exact nature of this 3-dimensional theory in this paper, especially we won't discuss anything about gravity, so we avoid such words to prevent possible misunderstandings.) Discussions of such kind of dualities of 3-dimensional Chern-Simons theory already exist in physics literature. Although we don't know how to define such a 3-dimensional arithmetic gauge theory, we will provide several pieces of evidence to show that this is something plausible: we will see that its relation with our 'current group' on number fields, closely mimics the relation between 3-dimensional Chern-Simons theory and 2-dimensional current algebra, as discussed in \cite{Witten2}. Also, we will see that one can get natural physical interpretations of several important but somewhat mysterious ingredients of the 2-dimensional theory from the point of view of this 3-dimensional theory, which is hard to see from the 2-dimensional theory itself.   

In section 2, we start from describing in certain detail some background and references of our ideas. We will take analogues of things we already know in order to get hints, and we make things precise whenever we touches number theory. Then we show that the S-duality of this 3-dimensional theory reflected in the path integral of our two dimensional theory, gives us the Gauss quadratic reciprocity law. Things will become rather concrete when we actually get to the quadratic reciprocity law, despite that the physics picture is quite conjectural. We will indicate possible generalizations of this idea aiming to give the same sort of physical interpretations of some higher power reciprocity laws, but we also point out some technical difficulties. Our discussion on this topic is hypothetical, and of course preliminary at best. However, along the way, we will see how some basic but intricate algebraic number theory come up from physical considerations, and furthermore a lot subtle ingredients of quantum field theory and number theory mingle together. The best hope is that our attempt may lead to a framework of providing physical interpretations of number theory reciprocity laws from which physicists will find it easier to understand algebraic number theory, and we may be able to make new conjectures in number theory from the framework.

\section{A physical Interpretation of Tate's Thesis}
For introductory material on Tate's thesis, one can see for example \cite{Kudla}.

In \cite{Witten}, Witten formulated several quantum field theories on an (smooth, complete) algebraic curve over an algebraically closed field. Here we will try to formulate a simplest possible conformal field theory on an algebraic number field from a somewhat different point of view. We will use some ideas of \cite{Witten}, of course, especially we will take some analogues of these ideas to apply to the case of number fields for guidance. We have no intention to make our discussion here rigorous or complete, however. Our goal here is to tentatively explore this possible connection between number theory and physics. We will see that much of Tate's thesis come out from physical considerations.

Let $K$ be a number field, and $\textsl{O}_K$ its ring of integers. Let $A_K$ be the ring of adeles, and $I_K$ the idele group, $C_K$ the idele class group, and $\textsl{I}_K$ the ideal class group. We denote by $\tau$ the diagonal embedding of $K^{\times}$ into $I_K$. We fix a global additive character $\psi$ of $A_K$, trivial on $K$. For any local embedding $F$ of $K$, let $d^{\times} x$ denote the multiplicatively invariant Haar measure on $F$ normalized so that the (local) units have volume $1$. Also we denote by $dx$ the self-dual additively invariant Haar measure with respect to the local component of $\psi$. By abuse of notation, we also denote by $d^{\times} x$ (and $dx$) the multiplicatively (and additively) invariant Haar measure on $I_K$ given by multiplying the local Haar measures. Now we will attempt to describe what one may call the $\operatorname{GL}(1)$ 'current group' on a number field.

First of all, for a commutative ring, we have at our hand the geometric object given by the prime spectrum of the ring, to be used to take analogue with the geometric case. For any place $v$ of $K$, local operators are in $\text{Hom}(\text{Spec} K_v, \operatorname{GL}(1))=\operatorname{GL}(1,K_v)$. By taking analogue of the discussion on multiplicative Ward identities in \cite{Witten}, if the local operator $f_v$ has negative valuation, then physically it corresponds to a positive energy excitation at $v$. So globally, quantum fields live in $\prod_{v}\operatorname{GL}(1,K_v)$, with the restriction that just like ordinary conformal field theory, for all but finitely many places $v$, $f_v$ lives in $\operatorname{GL}(1,O_v)$. So, in other words, quantum fields are elements of the idele group $I_K$.

Next, any two quantum fields differing by an element of $\tau(K^{\times})$ should be regarded as the same. We have reasons for imposing this requirement: one may consult section V of \cite{Witten}. Multiplying by elements of $\tau(K^{\times})$ is the analogue of conformal symmetry transformation.

So the path integral should be on the idele class group $C_K$. To integrate, we need a measure which should be an analogue of what physicists call the Feynman measure on the space of fields. In ordinary quantum field theory on flat spacetime, this (undefined) concept of Feynman measure should be translational invariant, which can be regarded as a consequence of the symmetries of flat spacetime. In our multiplicative case, the analogue of this is the requirement that the measure should be multiplicatively translational invariant. So this measure has to be the Haar measure on $C_K$ with an ambiguity of a scalar, which makes perfect sense.

Next, in path integral formulation of ordinary quantum mechanics and quantum field theory, expressions like $e^{iHt}$, $e^{i\int Ldt}$, or $e^{\int Ldx}$ show up essentially because of the Schrodinger equation, which itself can be regarded more or less as a consequence of the basic principles of quantum mechanics and the flat spacetime Lorentz symmetry. (There are many discussions on this issue, and we won't discuss it here. Note that the Schrodinger equation itself is not Lorentz invariant.) Here on the ideles, we have the multiplicative translational symmetry for the Haar measure, so what substitutes $e^{\int Ldx}$ in the path integral should be a multiplicative function on $C_K$ (Note that formally, $e^{iHt}$ is a quasicharacter on the additive group of $t$, which is a consequence of the Schrodinger equation.), which is nothing but $\omega\omega_s$ in general (on physics grounds, we assume that this function should be continuous. Furthermore we will provide physical interpretations of $\omega$ and $\omega_s$ in the next section), where $\omega$ is a Hecke character on $I_K$, and $\omega_s$ is the quasicharacter given by
\begin{equation}
\omega_s(x)=|x|^s
\end{equation}
for any $x\in I_K$. Where $s$ is a complex number.

Note that $\omega$ and $\omega_s$ can be factorized as products of local characters, and this is consistent with integrating the Lagrangian density over spacetime in ordinary quantum field theory (or over the worldsheet in two dimensional conformal field theory).\

Before we go any further, let us stop and make an observation which gives us a hint of why our construction possibly can come from a gauge theory. For an ordinary $U(1)$ gauge theory, the path integral should sum over all possible $U(1)$ principal bundles over the base manifold. Here, we have the canonical isomorphism
\begin{equation}\label{4.11}
\mathrm{Pic}(\texttt{Spec} \textsl{O}_K)\cong\textsl{I}_K
\end{equation}
Where the Picard group $\mathrm{Pic}(\texttt{Spec} \textsl{O}_K)$ classifies the isomorphism classes of invertible sheaves on $\texttt{Spec} \textsl{O}_K$. In fact, our path integral on $C_K$ somehow sums over $\textsl{I}_K$:\\
In number theory, there is a canonically defined surjective group homomorphism from the idele class group to the ideal class group: 
\begin{equation*}
\pi: C_K\rightarrow \textsl{I}_K
\end{equation*}
with
\begin{equation*}
Ker\pi=I(S_{\infty})/\tau(R^{\times})
\end{equation*}
where
\begin{equation*}
I(S_{\infty})=\prod_{\text{archimedean places}} K_v^{\times}\times \coprod_{\text{nonarchimedean places}} R_v^{\times}
\end{equation*}
where $R_v^{\times}$ is the group of (local) units in $\textsl{O}_v$, and $R^{\times}$ is the group of global units of $\textsl{O}_K$. So integration over $C_K$ already includes a summation over $\textsl{I}_K$. $\pi$ refines the information in the ideal class group, whose usefulness is illustrated by global class field theory. For us, it's usefulness is revealed by the path integral.

Before we can write down the path integral, we still have to consider the insertion of local operators. In ordinary quantum field theory, we have expressions like
\begin{equation}
\int \phi(x)e^{\int L(\phi(x))d^Dx}D\phi
\end{equation}
However, it is hard to make sense of it unless one makes the inserted operators have good decaying properties, and thinks of the measure as a linear map from some space of functions to $\mathbb{R}$. See \cite{Borcherds} for discussions on this issue.

To integrate over $C_K$, the integrand should be functions on $C_K$. Of course, the insertion of local operators should carry appropriate physical meaning. To think about what is the form our insertion should look like, here we consult the form of Polyakov path integral. See for example, \cite{Polchinski}, equation (3.5.5): for the inclusion of a particle, one inserts in the path integral a local vertex operator given by the state-operator correspondence. Furthermore, to make the vertex operator insertions diff-invariant, one integrates them over the worldsheet.

To mimic this process of insertion of vertex operators, we start from an unkown function $f(x)$ on $I_K$ which is a product of local functions with suitable decaying properties, and carrying appropriate physical meaning. Then we sum over $K^{\times}$ to make it $K^{\times}$ invariant (so we insist that $f(x)$ should make the following sum convergent):
$$
\sum_{\alpha\in K^{\times}} f(\alpha x)
$$
\begin{remark}
From the above, it is not correct to say that $K^{\times}$ should be the analogue of the string worldsheet, since the Polyakov path integral is intended to calculate string S-matrices, whereas our path integral is to be regarded as a path integral in conformal field theory. Rather, it makes some sense to regard $\texttt{Spec} \textsl{O}_K$ as the analogue of the worldsheet. But as we will see, archimedean places will also matter, so one should really say that the theory is to live on number fields.
\end{remark} 
Finally we can write down our path integral:
\begin{equation}
\int_{I_K/\tau(K^{\times})}\sum_{\alpha\in K^{\times}} f(\alpha x)(\omega\omega_s)(x)d^{\times}x
\end{equation}
Furthermore, note that $(\omega\omega_s)(\alpha x)=(\omega\omega_s)(x)$, for any $\alpha\in K^{\times}$. So the above equals
\begin{equation} 
\int_{I_K}f(x)(\omega\omega_s)(x)d^{\times}x
\end{equation}
which is exactly the global zeta integral $\textsl{z}(s,\omega;f)$ for the test function $f$. So we propose that the allowed functions should be in $S(A_K)$, the space of Schwartz-Bruhat functions on $A_K$.

Note that in the above integral, we have only one parameter $s$ which can be continuously varied. So it's tempting to regard $s$ as coming from the 'coupling constant'. We will see what this means as a coupling constant in the next section.

If we allow $f$ to vary, the global zeta integral $\textsl{z}(s,\omega)$ becomes a distribution, which is well known to be convergent for $\Re(s)>1$, and has a meromorphic analytic continuation to the whole $s$ plane and satisfies the functional equation
\begin{equation}\label{4.17}
\widehat{\textsl{z}(1-s,\omega^{-1})}=\textsl{z}(s,\omega)
\end{equation}
Where the global Fourier transform $\widehat{}$ is defined after we fix the additive character $\psi$ of $A_K$. We also have the functional equation for the complete global L function
\begin{equation} 
\Lambda(s,\omega)=\epsilon(s,\omega)\Lambda(1-s,\omega^{-1})
\end{equation}
which is independent of the choice of $\psi$.

\eqref{4.17} tells us that we can use analytic continuation to define our quantum theory for any value of the coupling constant $s$. We will make use of this fact in the next section discussing reciprocity laws. 
\begin{remark}
If we switch from number fields to global function fields (namely, function fields over a finite field), since Tate's thesis works for both cases, all the above discussion is essentially valid, except that we don't need to worry any more about archimedean places, and also we don't need to take analogues between number fields and function fields. (For the global function field case, we also have a canonical group homomorphism from the idele class group to the divisor class group, which should replace our discussion above around \eqref{4.11}. ) It is interesting to note that \cite{Witten} discusses quantum field theories on curves over an algebraically closed field, where Witten uses algebraic constructions relying on the algebraically closedness of the ground field, and he also remarks: " While one would wish to have an analogue of Lagrangians and quantization of Lagrangians in this more general setting, such notions appear rather distant at present." On the other hand, if we apply our discussion to global function field case, we are actually discussing conformal field theory on curves over a finite field. What we were trying to do, was just to write down a path integral which mimics a path integral in ordinary quantum field theory. But our discussion is not valid for curves over algebraically closed fields.
\end{remark}
\section{Toward Gauss reciprocity law and beyond}
First of all, let us briefly recall the relation between 3 dimensional Chern-Simons theory and 2 dimensional current algebra as discussed in \cite{Witten2}: 

For the simplest case without Wilson loops, in order to solve the quantum Yang-Mills theory with Chern-Simons action on an arbitrary three manifold $M$, we first chop $M$ into pieces, then solve the problem on the pieces, and then glue things back together. On a piece $\Sigma \times \mathbb{R}$, where $\Sigma$ is a closed surface, quantization of the theory is tractable by canonical quantization. With a certain gauge choice, solutions of the classical equation of motion gives us a finite dimensional phase space $\textit{M}$, the moduli space of flat connections on $\Sigma$ modulo gauge transformations. One knows that $\textit{M}$ has finite volume with respect to its natural symplectic structure, and in particular this implies that the quantum Hilbert space is finite dimensional. On the other hand, to actually get the quantum Hilbert space, one may first pick a complex structure $J$ on $\Sigma$, together with a linear representation of the gauge group $G$. For $G=SU(N)$ together with its fundamental representation, the moduli space $\textit{M}$, written as $\textit{M}_J$, can be reinterpreted as the moduli space of all stable rank $N$ holomorphic vector bundles of vanishing first chern class. $\textit{M}_J$ is a complex Kahler projective variety, and the quantum Hilbert space is the space of global holomorphic sections of certain line bundle on $\textit{M}_J$. Furthermore, one has a prescription to get rid of the choice of $J$ by a canonical flat connection on certain vector bundles on moduli space of $J$. 

On the other hand, if one considers current algebra on a Riemann surface, with a symmetry group $G$ at some level, then the Ward identities uniquely determine the correlation functions for descendants of the identity operator in genus zero case. However, if the genus of the Riemann surface is greater than zero, then in general the space of solutions of the Ward identities for descendants of the identity is a finite dimensional vector space called the 'space of conformal blocks', which is the same as the quantum Hilbert space obtained by quantizing the 3 dimensional theory as recalled above, as is shown by works of Segal and Witten. Witten remarked that this is the secret of the relation between these two theories!

To make analogies of these for number fields, we first recall some work on formal analogies between number fields and three manifolds started with the work of B. Mazur and others. $\texttt{Spec} \textsl{O}_K$ should be regarded as at least 3-dimensional from the point of view of etale cohomology. In \cite{Ma}, it is shown that the etale cohomology groups $\textsl{H}_{et}^n(\texttt{Spec} \textsl{O}_K,\mathrm{G}_m)$ vanish for $n>3$, and they are equal to $\mathbb{Q} / \mathbb{Z}$ for $n=3$. Furthermore these cohomology groups satisfy Artin-Verdier duality which is reminiscent of 3-dimensional Poincare duality. From these and other evidences people are suggesting analogies between $\texttt{Spec} \textsl{O}_K$ and three manifolds. Points in $\texttt{Spec} \textsl{O}_K$ (prime ideals) can be viewed as 1-dimensional objects and are compared to knots in a 3-manifold. In particular the absolute Galois group of a finite field is isomorphic to the profinite completion of $\mathbb{Z}$, the Fundamental group of a circle.

Having said all these, our postulate is that our 2-dimensional theory originates from a 3-dimensional gauge theory on $\texttt{Spec} \textsl{O}_K$ from the point of view of etale topology, which is the arithmetic counterpart of the 3-dimensional Chern-Simons theory with $G=U(1)$ with some sort of S-duality, as we have said. In the following, we will provide evidences for our postulation, and provide explanations of some mysteriously looking ingredients (for example the unexplained origin of the quasicharacter in section 1) in the 2-dimensional path integral from considerations of the 3-dimensional theory.

Let's first look at what the 'classical phase space' should be of such a 3-dimensional gauge theory. The moduli space of gauge equivalence classes of flat connections on $\Sigma$ corresponds to equivalence classes of homomorphisms
\begin{equation}
\phi: \pi_1(\Sigma)\rightarrow G
\end{equation}
, up to conjugation. 
Obviously, the arithmetic counterpart of $\pi_1(\Sigma)$ is the Galois group of the maximal unramified extension of $K$. With $G=U(1)$, abelian characters of this Galois group factors through the quotient by its commutator, which is the Galois group of the maximal abelian unramified extension of $K$, and is isomorphic to the ideal class group of $K$ by Hilbert class field theory. So, the classical phase space of our 3-dimensional gauge theory as a group should be isomorphic to the dual group of the ideal class group $\textsl{I}_K$ of $K$. So heuristically, the finiteness of the class number of an algebraic number field, seems to be the arithmetic counterpart of the finite volume property of $\textit{M}$, or the finite dimensional property of the quantum Hilbert space of 3d Chern-Simons on $\Sigma \times \mathbb{R}$. The finiteness of the dimension of the latter space in presence of certain Wilson operator insertions is closely related with the skein relations in knot theory as revealed by section 4 of \cite{Witten2}. 

Let's look at the same phase space from another point of view. As we recalled above, upon picking a complex structure on $\Sigma$, for $G=U(1)$, the phase space has another interpretation as the moduli space of stable holomorphic line bundles on $\Sigma$ with vanishing first chern class. The analogue for this in our setting is the finite Picard group $\mathrm{Pic}(\texttt{Spec} \textsl{O}_K)$, which is isomorphic to the ideal class group $\textsl{I}_K$ for $K$. So either way, by taking formal analogues, we see that the classical phase space can be identified as a group with the ideal class group.

Quantization of 3-dimensional Chern-Simons theory on $\Sigma \times \mathbb{R}$, as we recalled, is by taking the space of global holomorphic sections of some line bundle on $\textit{M}_J$. Here $\textit{M}_J$ is substituted by the finite group $\textsl{I}_K$, so the quantum Hilbert space is just a finite dimensional complex vector space with dimension equal to the class number of $K$. 

Next, let us consider what should be the 'space of conformal blocks' for our theory of current group on $K$. Since we have a path integral, we can easily get the quantum equation of motion, and Ward identities for symmetries by certain change of variables in the path integral (for introductory material on this topic, the reader can refer to chapter 9 of \cite{Peskin}):
For any idele $\alpha\in I_K$, we make a change of variables $x\rightarrow \alpha x$ in the path integral, the translational invariance of the Haar measure tells us that
\begin{equation}\label{2.2}
\int_{I_K}f(x)(\omega\omega_s)(x)d^{\times}x=\int_{I_K}f(\alpha x)(\omega\omega_s)(\alpha x)d^{\times}\alpha x=(\omega\omega_s)(\alpha)\int_{I_K}f(\alpha x)(\omega\omega_s)(x)d^{\times}x
\end{equation}
This should be regarded as the quantum equation of motion, and the (multiplicative) Ward identity in our setting is the same equation with the restriction that $\alpha\in\tau(K^{\times})$.

To get the description of the space of conformal blocks, we need to look at the class number from the point of view of the canonical homomorphism $\phi$ from the idele group to the ideal group as follows:
We consider an equivalence relation on the ideles $I_K$ defined as: two elements $v_1$ and $v_2$ are said to be equivalent if and only if valuations of them at every nonarchimedean place are equal. In other words, if they are mapped to the same element in the ideal group by $\phi$. The group $\tau(K^{\times})$ acts on the quotient $I_K/Ker(\phi)$ by 'pointwise' multiplications. The number of orbits in $I_K/Ker(\phi)$ with respect to this action is equal to the class number of $K$.
In ordinary conformal field theory, we need to study to what extent the Ward identities determine correlation functions for all descendants of the unit, inserted at any allowed combinations of points on the Riemann surface. Now if our Hecke character $\omega$ is trivial, then the counterpart of all descendants of the unit inserted at any allowed combinations of points should be the set of all possible $f(x)$, which are (restricted) products of local characteristic functions for some $\pi_p^{s}\textsl{O}_v$, where $s$ is the prescribed valuation at $p$. (Since it's reasonable to  say that local excitations of our quantum field are measured by local evaluations, as we have said in the previous section. Moreover this point will become clearer when we discuss insertions of Wilson loop operators.) By examining the Ward identity in our setting, it is clear that the dimension of the complex vector space of conformal blocks should be the number of orbits for the action of $\tau(K^{\times})$ on this set, where the action is given by
\begin{equation}
\alpha\in\tau(K)^{\times}\rightarrow (f(x)\rightarrow f(\alpha x))
\end{equation}
This action is the same as the action of $\tau(K^{\times})$ on the quotient $I_K/Ker(\phi)$ as we just discussed, so the number of orbits equals the class number of $K$. In other words, the complex vector space of conformal blocks for our theory of current group on $K$ for a trivial Hecke character, has dimension equal to the class number of $K$, which, as we have seen above, is also equal to the dimension of the quantum Hilbert space of our hypothetical 3-dimensional gauge theory. This mimics nicely the 'secret' of the relation between 3-dimensional Chern-Simons theory and 2-dimensional current algebra as we recalled! Furthermore, the condition that $\omega$ being trivial has the counterpart that there being no insertion of Wilson loops in 3d Chern-Simons theory (recall that all the above discussion of formal similarities is in the absence of Wilson loops). Being succeeded at this stage, let us next consider possible insertions of Wilson or t'Hooft operators in our theory to get something more interesting.

In \cite{Witten3}, Kapustin and Witten considered certain topological Wilson-t'Hooft operators in some 4 dimensional supersymmetric topological Yang-Mills theories reduced to two dimensions, and interpreted geometric Langlands as coming from the S-duality of the underlying 4 dimensional gauge theory switching Wilson and t'Hooft operators. Note that although we described our current group on a number field by the global zeta integral in section 1, we know very little about how to choose the test function $f(x)$, and know nothing about how to choose the quasicharacter. In the following, we will argue that to talk about charges, we need to make a choice of $s$, and insertion of certain t'Hooft operators in the hypothetical 3-dimensional theory should reflect itself in the 2-dimensional theory as insertion of certain Hecke characters in the path integral (to be more precise, our following arguments on this point for number fields other than $\mathbb{Q}$ are incomplete for some reasons, as we will see. But we expect some refinements will work.), and insertion of certain Wilson operators in the hypothetical 3-dimensional theory should reflect itself in the 2-dimensional theory as certain changes of $f(x)$. Then, the quadratic reciprocity law comes to surface if we switch the Wilson and t'Hooft operators by the hypothetical S-duality of the 3-dimensional theory! 

In the following we first restrict ourselves to quadratic reciprocity, and the discussions will be carried out on $K=\mathbb{Q}$. We will first see how the simplest case of quadratic reciprocity comes out as quickly as we can, then we refine our discussion to get the full quadratic reciprocity law. After that, we will say a bit about our ideas for algebraic number fields other than $\mathbb{Q}$.

In ordinary quantum field theory, the effect of including a Wilson loop operator in the path integral is to add an external charge in a certain representation of the gauge group, whose trajectory in spacetime is the loop; and the effect of including a t'Hooft operator is to instruct a certain singularity of the fields localized along the support of the t'Hooft operator. From the point of view of etale topology, we are interested in Wilson and t'Hooft operators supported along 'circles' $\texttt{Spec} \mathbb{F}_q$, where $q$ is a prime. So in the 2-dimensional theory, the effect of these operators are just insertions of some local operators. 

We first consider about t'Hooft operators supported on $\texttt{Spec} \mathbb{F}_q$. To detect the nature of a singularity of a connection field like the Dirac magnetic monopole, one can trace the field along loops in space where there is no singularity, and look at the resulting holonomy. From this point of view, we know how to describe the effect of a t'Hooft operator supported at $\texttt{Spec} \mathbb{F}_q$ in our settings: we need to take a covering of $\texttt{Spec}\mathbb{Z}$, or more precisely, a field extension of $\mathbb{Q}$, which is only ramified at $q$, and look at the 'monodromy' of other circles $\texttt{Spec} \mathbb{F}_p$ for primes $p\neq q$. Since here we are only interested in the quadratic reciprocity, we restrict ourselves to the case of a double covering, thus a quadratic extension. Note that when $q\equiv 1(\texttt{mod} 4)$, there is such a field extension $\mathbb{Q}\rightarrow\mathbb{Q}(\sqrt{q})$. We notice that very similar mathematical question has been considered in the very interesting article \cite{Kap}, where it is shown that the monodromy should correspond to the Frobenius element of the Galois group at $p$. (Also the authors interpret the Legendre symbols as linking numbers, and the Gauss reciprocity law as the interchanging symmetry of linking numbers.) For the case of a quadratic extension, this monodromy only depend on whether $p$ splits or not. Elementary number theory then tells us that the monodromy of an odd prime $p$ is given by $(\frac{q}{p})$, Where $(\frac{}{})$ denotes the Legendre symbol. While for $p=2$ (so $q\neq 2$), it's given by $\frac{q^2-1}{8}$, which equals the Kronecker symbol $(\frac{D}{2})$. (Kronecker symbol at the place $2$ is not multiplicative, however it is multiplicative on $1+4\mathbb{Z}_2$. Moreover it's the same as the Legendre symbol at any other places. Roughly speaking, the specialness of $2$ comes from the fact that we are considering quadratic symbols, so $2$ is the only special prime where the effect of Hensel's lemma for quadratic polynomials is different from the case of other primes. But the physics picture of the monodromy is the same anyway.) Then, how do we incorporate this into our 2-dimensional path integral? To simply get the correct monodromy, we propose that the effect of the inclusion of such a t'Hooft operator in the 2-dimensional path integral, is to add a function $\omega_q$ defined on ideles with component at $q$ equaling to $1$ as
\begin{equation}
\omega_q(x)=\prod_{\texttt{nonarchimedean places }p\neq q} (\frac{q}{p})^{-V_p(x)}
\end{equation}
Where we need the power $-V_p(x)$ because we need $\omega_q$ to be multiplicative in our multiplicative theory. Moreover, because of conformal symmetry, we need to require the function $\omega_q$ to come from a function on the idele class group. In other words, we require the $\tau(K^{\times})$ invariance of $\omega_q$. Once this restriction is put, $\omega_q$ is now a multiplicative function defined on all the ideles, trivial on $\tau(K^{\times})$, and thus is a Hecke character for which the conductor we do not know a priori without assuming quadratic reciprocity law. (There is a slight cheating to directly say that $\omega_q$ is a Hecke character, since we haven't shown that $\omega_q$ is continuous with respect to the topology of ideles. Of course if we are allowed to use the quadratic reciprocity law, there is no problem of showing this. But the point here is that we want to interpret quadratic reciprocity law by physics without first assuming it. On physics grounds, we say that $\omega_q$ is a Hecke character by pretending that it is continuous.)

Next let's decide how to include a Wilson loop operator at $p$. First of all, to talk about charges, we need to choose a (one dimensional) continuous representation of $\operatorname{GL}(1, A_K)$ which is a product of local representations, and restricts to the trivial representation on $\tau(K^{\times})$(because of conformal symmetry of the 2-dimensional theory). In other words, we need to choose a quasicharacter of the idele group whose restriction to $\tau(K^{\times})$ is $1$. Furthermore, it's unnatural for this quasicharacter to have any nontrivial conductor a priori (and in fact, as we have seen in the above, the insertion of t'Hooft operators secretly takes care of such a choice). So, what we need to choose is exactly an $\omega_s$, which is uniquely determined by the complex number $s$. It is also for this reason, that we may regard $s$ as the $\operatorname{GL}(1)$ coupling constant. (In fact, we will see that the choice of $s$ has zero total effect for the insertion of Wilson loop operators, once they are correctly done. But it does have crucial role in the realization of S-duality in our 2-dimensional path integral, because of \eqref{4.17}.) Once the representation is chosen, we can decide how to insert a Wilson loop operator at $p$: it instructs us to evaluate the monodromy at the 'loop' $p$ in the 2-dimensional path integral, in our chosen representation. In the 2-dimensional path integral, to include such an effect of a Wilson loop operator at $p$, the procedure then is to change $f(x)$ to $f(\alpha_p x)$, where $\alpha_p=(p,1,1,...,p,1,1,...)$ is an idele with norm one such that the nonarchimedean valuation of $\alpha_p$ is equal to $1$ at $p$, and $0$ at any other primes. Also, we should require positive valuation at the real place, since $-1$ at the real place may have nontrivial monodromy. Furthermore, We need to require the components of $\alpha_p$ to be $1$ at any other nonarchimedean place, and to be just $p$ at $p$, to avoid complications coming from unknown Hecke characters, or t'Hooft operator insertions. Then, it is clear that $\alpha_p$ is the unique idele satisfying all these restrictions. (Let's explain the norm one requirement: the reason of this is the same as it is explained in \cite{Witten2}: the total charge of a closed universe should be zero, since the electric flux has nowhere to go. So we need to require $\alpha_p$ to have norm $1$ in order that the representation $\omega_s$ takes it to $1$. Later we will discuss insertions of multiple Wilson loop operators. For that case, it suffices that the norm of the product of corresponding idele equals one, since only the total charge should be zero.)

Let's denote the Wilson operator at $p$ as $w_p$, and use a subscript $q$ to indicate the inclusion of a t'Hooft operator at $q$, or in other words, we insert a $\omega_q$ in the path integral. Let's evaluate the 'amplitude' $<w_p>_q$ for the Wilson loop at $p$ in the presence of a t'Hooft operator at $q$. According to the usual formula of quantum field theory, we have
\begin{equation}\label{2.5}
<w_p>_q=\frac{\int_{I_K}f(\alpha_p x)(\omega_q\omega_s)(x)d^{\times}x}{\int_{I_K}f(x)(\omega_q\omega_s)(x)d^{\times}x}
\end{equation}
Where $f(x)$ is a function to be inserted for the vacuum in the presence of t'Hooft operator insertions given by $\omega_q$. But we will see that we can get quadratic reciprocity without any knowledge of $f(x)$ other than that the denominator of the right hand side of the above equality is not zero.

By using the quantum equation of motion \eqref{2.2}, \eqref{2.5} gives us
\begin{equation}\label{6.100}
<w_p>_q=\omega_q(p)^{-1}=(\frac{q}{p})
\end{equation}
We see that this expression is independent of $s$. From this one sees an effect of the norm one requirement.

To get to special cases of quadratic reciprocity as quickly as possible, we first require that $p\equiv 1 (\texttt{mod }4)$, in addition to the requirement $q\equiv 1 (\texttt{mod }4)$. Then, apply the hypothetical S-duality of the 3-dimensional theory, the Wilson and t'Hooft operators get switched, in other words $\omega_q$ and $\alpha_p$ are transformed into $\omega_p$ and $\alpha_q$. After S-duality, we are able to describe the 2-dimensional physics in the same way, as we have seen in \eqref{4.17}, but we need to replace $\omega_p$ by $\omega_p^{-1}$. Furthermore, the Fourier transform has the effect of transforming the idele $\alpha_q$ into its inverse: to be more explicit, let $f_1(x)=f(\alpha_q x)$, we have
\begin{align*}\label{*}
\widehat{f_1(x)}\tag{*}
&=\int_{A_K} f_1(y)\psi(xy)dy\\
&=\int_{A_K} f(\alpha_q y)\psi(xy)dy\\
&=\int_{A_K} f(y_1)\psi(x\alpha_q^{-1}y_1)dy_1\\
&=\widehat{f}(\alpha_q^{-1}x)
\end{align*}
One notices that the fact that $\alpha_q$ has norm $1$ is crucial in the derivation of the above.
So from the 2-dimensional point of view, S-duality tells us that
\begin{equation}
<w_{q^{-1}}>_{p^{-1}}=<w_p>_q
\end{equation}
By \eqref{6.100}, the above equality is
\begin{equation}
(\frac{p}{q})=(\frac{q}{p})
\end{equation}
which is nothing but the Gauss quadratic reciprocity law for $p$ and $q$ for the special case when both $p$ and $q$ are congruent to $1$ mod $4$ ! 

Next, we refine the above discussion, and consider the insertion of t'Hooft and Wilson operators in more detail in order to get the full quadratic reciprocity law. If we intend to insert a t'Hooft operator at $q$ for an odd prime $q\equiv 3(\texttt{mod }4)$, then we are in trouble since there is no quadratic extension of $\mathbb{Q}$ that only ramifies at $q$. (We need also to take into account the ramifications of the field extension at archimedean places.) So in fact, if $q$ is congruent to $3$ mod $4$, to insert a t'Hooft operator at $q$, we have to secretly include some other insertions of the t'Hooft operator at some other places. As an example let's consider the field extension $\mathbb{Q}\rightarrow\mathbb{Q}(\sqrt{(-1)^{\frac{q-1}{2}}q})=L$, for the case when $q$ is an odd prime. When $q\equiv 1(\texttt{mod }4)$, this is the old good extension which only ramifies at $q$. But when $q\equiv 3(\texttt{mod }4)$, it also ramifies at infinity. So in the latter case, we should regard this field extension as giving us two t'Hooft operators inserted at $q$ and infinity, respectively. As another example we consider the field extension $\mathbb{Q}\rightarrow\mathbb{Q}(\sqrt{q})=L_1$ for the case when $q\equiv 3(\texttt{mod }4)$. This extension ramifies at $q$ and $2$, with ramification indexes both equal to $2$. However, there is something different at places $q$ and $2$: the discriminant $D=4q$, and so the exponents of $q$ and $2$ in the discriminant are $1$ and $2$, respectively. (For latter purpose, this can be equivalently stated as: considering the different of the field extension $L_1/\mathbb{Q}$, the differential exponent of the prime above $q$ is $1$, and of the prime above $2$ is $2$.) We need to include the information of these exponents as the multiplicities of the local insertion of t'Hooft operators. (This should be clear when we think of the algebraic-geometric picture of the effective different divisor, where differential exponent is interpreted as multiplicity in a definite way. From a differential-geometric point of view, this multiplicity corresponds to the first Chern class of a $U(1)$ bundle representing the field singularity.) So for the field extension $L_1/\mathbb{Q}$, we have inserted a t'Hooft operator at $q$ with multiplicity $1$, and a t'Hooft operator at $2$ with multiplicity $2$. If $q=2$, we can take the extension $\mathbb{Q}\rightarrow\mathbb{Q}(\sqrt{2})=L_2$. Then we get a single t'Hooft operator insertion at $2$ with multiplicity $3$. In any of these cases, elementary number theory again tells us that the combined monodromy of a prime $p$ which is unramified for a quadratic field extension is given by $(\frac{D}{p})$, Where $(\frac{}{})$ denotes the Kronecker symbol, and $D$ is the discriminant of the field extension. Again, to simply get the correct monodromy, the effect of the inclusion of a set of such t'Hooft operators (We don't need to specify the order of the insertions of t'Hooft operators, since we are considering an abelian theory.) given by a quadratic field extension $M/\mathbb{Q}$ in the 2-dimensional path integral, is to add to the path integral a multiplicative function $\omega_{M/\mathbb{Q}}$ defined on ideles as: 
For ideles whose components equal $1$ at all ramified places 
\begin{equation}
\omega_{M/\mathbb{Q}}=\prod_{\texttt{nonarchimedean places }p \texttt{ prime to }D} (\frac{D}{p})^{-V_p(x)}
\end{equation}
One may wonder why we don't include the possible monodromy coming from the real place, as for the place $p=\infty$, the monodromy is $-1$ if the local field extension is ramified at $\infty$, and otherwise it is equal to $1$. The reason we can ignore this issue is that we forbid the coexistence of Wilson and t'Hooft operators at the same place, which excludes the possibility for a nontrivial monodromy coming from the real place. Next we should require that $\omega_{M/\mathbb{Q}}$ becomes trivial when restricted to $\tau(K^{\times})$, as before. Again, on physics grounds, we say that $\omega_q$ is a Hecke character by pretending that it is continuous. (Actually one can show that it is continuous if one is allowed to use the quadratic reciprocity law.)

Next we consider the insertion of Wilson operators in more details. First note that since we have to include in the theory insertion of t'Hooft operators at infinite primes, by S-duality we also have to think of the problem of insertion of Wilson loop operators at these primes (even though they won't have observable effects in the 2-dimensional path integral for the reason stated above, we still have to include them as they are required by S-duality). We can straightforwardly extend our discussion for the nonarchimedean cases, and see that Wilson loop operators inserted at the real place has the effect of changing $f(x)$ to $f(\alpha_{\infty}x)$, where the idele $\alpha_{\infty}=(-1,1,1,1,...)$, with the $-1$ at the real place. Again, this idele is the unique idele satisfying all our previous requirements. Next, we need to take into account the multiplicity of a local insertion of Wilson operator at $p$, again as required by S-duality. This is easy: since we are working with a multiplicative theory, we just need to raise the idele $\alpha_p$ to the power given by the multiplicity $m$, and make change to $f(x)$ using this new idele: $f(x)\rightarrow f(\alpha_p^m x)$. This has the effect of raising the monodromy to the power given by the multiplicity. Also, as we mentioned before, to satisfy the total charge zero requirement, one has to make sure that the norm of the products of all ideles $\alpha_p^m$ is equal to $1$. Then, one has obvious generalization of equation (*), which tells us how to operate with S-duality transformation. In practice, it makes no difference if we further require that the norm of each $\alpha_p$ is equal to $1$ (Sometimes it makes things easier to state). Lastly, the insertion of Wilson operators should always come in a set with appropriate multiplicities in order that after S-duality transformation, they give a legal set of insertions of t'Hooft operators.

Now we apply S-duality to several different cases to get the quadratic reciprocity law for primes $p$ and $q$ that are different from each other:

First of all, it's obvious that $(\frac{-1}{q})=-1$ if $q\equiv 3(\texttt{mod }4)$. This is the most trivial case for which we don't actually need a physical interpretation. (and we will see that this is the only case of quadratic reciprocity, for which we don't have a physical interpretation from S-duality discussed above!)

Next we determine $(\frac{-1}{p})$ for $p\equiv 1(\texttt{mod }4)$. We consider t'Hooft operator insertions given by the field extension $\mathbb{Q}\rightarrow\mathbb{Q}(\sqrt{-1})$, and an insertion of Wilson operator at $p$ with multiplicity $1$. Then as before, after applying S-duality, we get Wilson operators at $\infty$ and $2$ with multiplicities $1$ and $2$, respectively. Also we have a t'Hooft operator at $p$ with multiplicity $1$. So the path integral gives
\begin{equation}
(\frac{-4}{p})=(\frac{p}{2})^2
\end{equation}
This equality tells us that $(\frac{-1}{p})=1$.
 
For the case when both $p$ and $q$ are odd, and at least one of them is congruent to $1$ mod $4$, say $p\equiv 1(\texttt{mod }4)$, we consider the insertion of a Wilson operator at $p$ with multiplicity $1$, and t'Hooft operator insertions given by $L/\mathbb{Q}$. Then after applying S-duality, we get Wilson operators at $q$ and $\infty$ both with multiplicity $1$, and a t'Hooft operator at $p$ with multiplicity $1$. S-duality transformation of the path integral gives
\begin{equation}
(\frac{D}{p})=(\frac{p}{q})
\end{equation}
This is exactly the quadratic reciprocity law for $p$ and $q$ in this case.

Next, consider the case when $q$ is an odd prime, and $p=2$. We consider t'Hooft operator insertions given by $L/\mathbb{Q}$, and a single Wilson operator insertion at $2$ with multiplicity $3$. Then as before, S-duality gives
\begin{equation}
(\frac{D}{2})^3=(\frac{8}{q})
\end{equation}
which obviously reduces to the quadratic reciprocity law for $2$ and $q$.

Finally, we consider the case when both $p$ and $q$ are odd primes congruent to $3$ mod $4$. This time, t'Hooft operator insertions are given by $L/\mathbb{Q}$, and there are Wilson operators at $p$ and $2$ with multiplicities $1$ and $2$, respectively. So S-duality gives
\begin{equation}
(\frac{D}{p})(\frac{D}{2})^2=(\frac{4p}{q}) 
\end{equation}
again this equality reduces to the quadratic reciprocity law for $p$ and $q$ in this case.
\begin{remark}
Note that the specific choice of $f(x)$ doesn't matter in this story.
\end{remark}

Next, we want to try to generalize our discussion to an arbitrary algebraic number field $K$, with the hope of getting more reciprocity laws. However, we will see that although Wilson and t'Hooft operator insertions seem to generalize without difficulty, there are other difficulties one should overcome before one can get any higher power reciprocity laws. 

For a number field $K$, we fix a uniformizer $\pi_p$ at each place $p$. An insertion of a set of t'Hooft operators is given by a finite abelian extension $L/K$ of a prime degree (for simplicity). As usual, we denote by $\mathfrak{\delta}_{L/K}$ the different of the field extension, and by $\mathfrak{D}_{L/K}$ the relative discriminant. Let $\mathfrak{R}_{L/K}$ denote the finite set of places of $K$ that are ramified in this extension. Then we have t'Hooft operator insertions at places $q$ in $\mathfrak{R}_{L/K}$ with multiplicities given by the differential exponent of any prime above $q$ (this is equal to the exponent of $q$ in $\mathfrak{D}_{L/K}$). Furthermore, the monodromy of an unramified loop (place) $p$ should be given by the local Frobenious element, or the local norm residue symbol $(\frac{\pi_p, L/K}{p})$. (This is a well defined element in the Galois group since $L/K$ is abelian.) A Wilson operator with multiplicity $m$ at a place $p$ is to change $f(x)$ to $f(\alpha_p^m x)$, where $\alpha_p$ is any idele given by: if $p$ is nonarchimedean, then the component at any nonarchimedean place other than $p$ is $1$, and the component at $p$ is $\pi_p$, the component at any real place is positive, and the norm equals $1$; if $p$ is a real place, then the component at $p$ equals $-1$, and at any nonarchimedean place equals $1$, and has positive valuation at any other archimedean place, and the norm equals $1$. If $p$ is a complex place, $\alpha_p$ has component $1$ at any nonarchimedean place, and has positive valuation at any archimedean place, and has norm $1$. Note that $\alpha_p$ is not uniquely determined at archimedean places, but it's obvious that any two choices of $\alpha_p$ will have exactly the same effect in the path integral. Furthermore, from the above description it is easy to see that Wilson operators at complex places have no effect at all in the path integral, so we can just instead put the restriction that there are no insertion of Wilson operators at complex places. In fact this is required by S-duality: since no complex place can possibly be ramified, no t'Hooft operators can be inserted at complex places, so S-duality tells us no Wilson operators can be inserted at complex places. Also, we require that no Wilson and t'Hooft operators can be inserted simultaneously at a single place.

Note that the monodromy lives in the Galois group. In order to discuss higher power reciprocity laws, we need to first decide how to take the values of monodromies living in different Galois groups in the 2-dimensional path integral. For quadratic extensions, there is only one way to identify the Galois group with the group $\mathbb{Z}/2\mathbb{Z}$, so we don't need to worry about different identifications. This is one of the reasons why quadratic reciprocity law is much easier to get. However, for higher order extensions, such naive unique identifications no longer exist. In the following, we will explain that an alternative idea also fails:

Kummer theory is one central ingredient for the definition of Hilbert symbols and higher reciprocity laws. One may try to use Hilbert symbols to define the values of monodromies, and hope to get some higher reciprocity laws. So next, we let $K$ to be the cyclotomic field $K=\mathbb{Q}(\zeta_{p_0})$, where $p_0$ is any fixed odd prime. For any Kummer extension $L/K$, Kummer theory and local class field theory combined gives us the local Hilbert symbol $(\pi_p, L/K)_p$ of order $p_0$, taking values in the group of $p_0$th roots of unity. Let's try to take this as the value of the monodromy detected by $p$ in the presence of the set of t'Hooft operator insertions given by $L/K$, and see if it works. Then as before, the t'Hooft operator insertions given by $L/K$, should have the effect in the 2-dimensional path integral of inserting a multiplicative function $\omega_{L/K}$ on the idele group given by
\begin{equation}
\omega_{L/K}(\alpha)=\prod_{p\notin \mathfrak{R}_{L/K}}(\pi_p, L/K)_p^{-v_p(\alpha)}
\end{equation}
For ideles $\alpha$ whose component at any place in $\mathfrak{R}_{L/K}$ is equal to $1$. The product in the above equation is well defined since for all but finitely many places, the local Hilbert symbol equals $1$. Again, we require that the restriction of $\omega_{L/K}$ to $\tau(K^{\times})$ is trivial, and it is straightforward to check that $\omega_{L/K}$ is uniquely defined. On physics grounds we pretend it is continuous, and so $\omega_{L/K}$ defines a Hecke character.

Now we pick two different prime elements $p$ and $q$ in the ring of integers $\mathbb{Z}[\zeta_{p_0}]$ of $K$, both being coprime to $p_0$. We consider two Kummer extensions $L=K(\sqrt[p_0]{p})/K$, and $M=K(\sqrt[p_0]{q})/K$. For the cyclic extension $K/\mathbb{Q}$, only $p_0$ ramifies, and we have $(p_0)=(1-\zeta_{P_0})^{p_0-1}$ as ideals in $K$. Let's denote the ideal $(1-\zeta_{p_0})$ by $v$. Kummer theory tells us that the only possible ramifications of $L/K$ are at $p$ and $v$, and the only possible ramifications of $M/K$ are at $q$ and $v$. We need to make the requirement that $v$ is unramified for at least one of $L/K$ or $M/K$, and under this requirement, we hope to see that the $p_0$th power reciprocity law for $p$ and $q$ follows from S-duality. 

Note that $p$ is tamely ramified for $L/K$ with exponent $p_0-1$ in the relative discriminant, and $q$ is tamely ramified for $M/K$ with exponent $q_0-1$ in the relative discriminant. Without loss of generality, we assume that $v$ is unramified for $L/K$. Then we consider t'Hooft operator insertions given by $M/K$, and an insertion of Wilson operator at $p$ with multiplicity $p_0-1$ at $p$. So there is a t'Hooft operator at $q$ with multiplicity $q_0-1$, and a possible t'Hooft operator at $v$ with multiplicity given by the exponent $\exp_{\mathfrak{D}_{M/K}}v$ of $v$ in the relative discriminant $\mathfrak{D}_{M/K}$. 

Therefore, the S-duality transformation of the 2-dimensional path integral should give us
\begin{equation}
(\pi_p,M/K)_p^{p_0-1}=(\pi_q,L/K)_q^{p_0-1}(1-\zeta_{p_0},L/K)^{\exp_{\mathfrak{D}_{M/K}}v}
\end{equation}
By the skew symmetry of the local Hilbert symbol, we know that the above is equivalent to
\begin{equation} \label{6.17}
(p,q)_p(p,q)_q=(1-\zeta_{p_0},L/K)^{-\exp_{\mathfrak{D}_{M/K}}v}
\end{equation}
Since $v$ is unramified for $L/K$, and $q$ is coprime to $p_0$, we have $(p,q)_v=(q,p)_v^{-1}=1$, Hilbert reciprocity law then tells us that if \eqref{6.17} is true, we should have
\begin{equation}
(1-\zeta_{p_0},L/K)^{\exp_{\mathfrak{D}_{M/K}}v}=1
\end{equation}
However, obviously it's possible choose certain $p$ such that $(1-\zeta_{p_0},L/K)$ is nontrivial. Furthermore we may choose $q$ such that $\exp_{\mathfrak{D}_{M/K}}v$ is not divisible by $p_0$. (An explicit example is: $q_0=3$, $q=5$, then $\exp_{\mathfrak{D}_{M/K}}v=4$, not divisible by $3$. One can demonstrate this by calculating the relative discriminant explicitly by using local and global conductors, and solving some explicit congruence equations. But we will skip the details since it seems irrelevant with our main discussion.) This means that our naive attempt to get the value of monodromy by using Hilbert symbols is failed.
\begin{remark}
Despite the difficulties, it seems reasonable to expect that, possibly with some more effort, one may get more reciprocity laws from this S-duality. Moreover exploration in this direction may also help clarify the unknown physics picture of the hypothetical 3-dimensional gauge theory if it does exist as expected. Although our discussion is hypothetical, one can see that a lot interesting and subtle ingredients in both quantum field theory and number theory show up and mingle together. At least, as we have seen, one does get the full quadratic reciprocity law from S-duality. So we hope that at least some of the ideas presented here will be of interest for further explorations.    
\end{remark}
\section*{Acknowledgments}
I would like to thank Richard Borcherds, Shenghao Sun, Chul-hee Lee, and Kenneth Ribet for useful discussions.

%\cite[ [1] Hartshorne, R., $Algebraic Geometry$, GTM 52.]
\bibliographystyle{amsplain}

\end{document}